\documentclass{sig-alternate}
\usepackage{amssymb,amsfonts,amsmath}
\usepackage{epsfig}
\usepackage{graphicx}
\usepackage{url}
\usepackage{multirow}
\usepackage{enumerate}

\usepackage[usenames,dvipsnames,svgnames,table]{xcolor}

\begin{document}

\title{Topic Clusters and Diversity of User Interests}
\title{Topical Diversity and Popularity in Social Media}
\title{Messages and Messengers: Topical Diversity and Social Impact}
\title{Topicality and Social Impact: \\Diverse Messages but Focused Messengers}
\numberofauthors{1}
\author{
\alignauthor Lilian Weng \quad Filippo Menczer \\
\affaddr{Center for Complex Networks and Systems Research} \\
\affaddr{School of Informatics and Computing, Indiana University Bloomington}
}
\maketitle

\begin{abstract}

Are users who comment on a variety of matters more likely to achieve high influence than those who delve into one focused field? Do general Twitter hashtags, such as \#lol, tend to be more popular than novel ones, such as \#instantlyinlove? Questions like these demand a way to detect topics hidden behind messages associated with an individual or a hashtag, and a gauge of similarity among these topics.
Here we develop such an approach to identify clusters of similar hashtags by detecting communities in the hashtag co-occurrence network. Then the topical diversity of a user's interests is quantified by the entropy of her hashtags across different topic clusters. A similar measure is applied to hashtags, based on co-occurring tags.
We find that high topical diversity of early adopters or co-occurring tags implies high future popularity of hashtags. In contrast, low diversity helps an individual accumulate social influence. In short, diverse messages and focused messengers are more likely to gain impact.

\end{abstract}

\category{}{Information systems}{Information systems applications}[Collaborative and social computing systems and tools, Social networking sites]
\category{}{Information systems}{World Wide Web}[Web applications, Web mining]
\category{}{Human-centered computing}{Collaborative and social computing}[Collaborative and social computing theory, concepts and paradigms, Social media]
\category{}{Applied computing}{Law, social and behavioral sciences}[Sociology]



\section{Introduction}

\begin{figure}[t!]
\centering
\includegraphics[width=0.75\columnwidth]{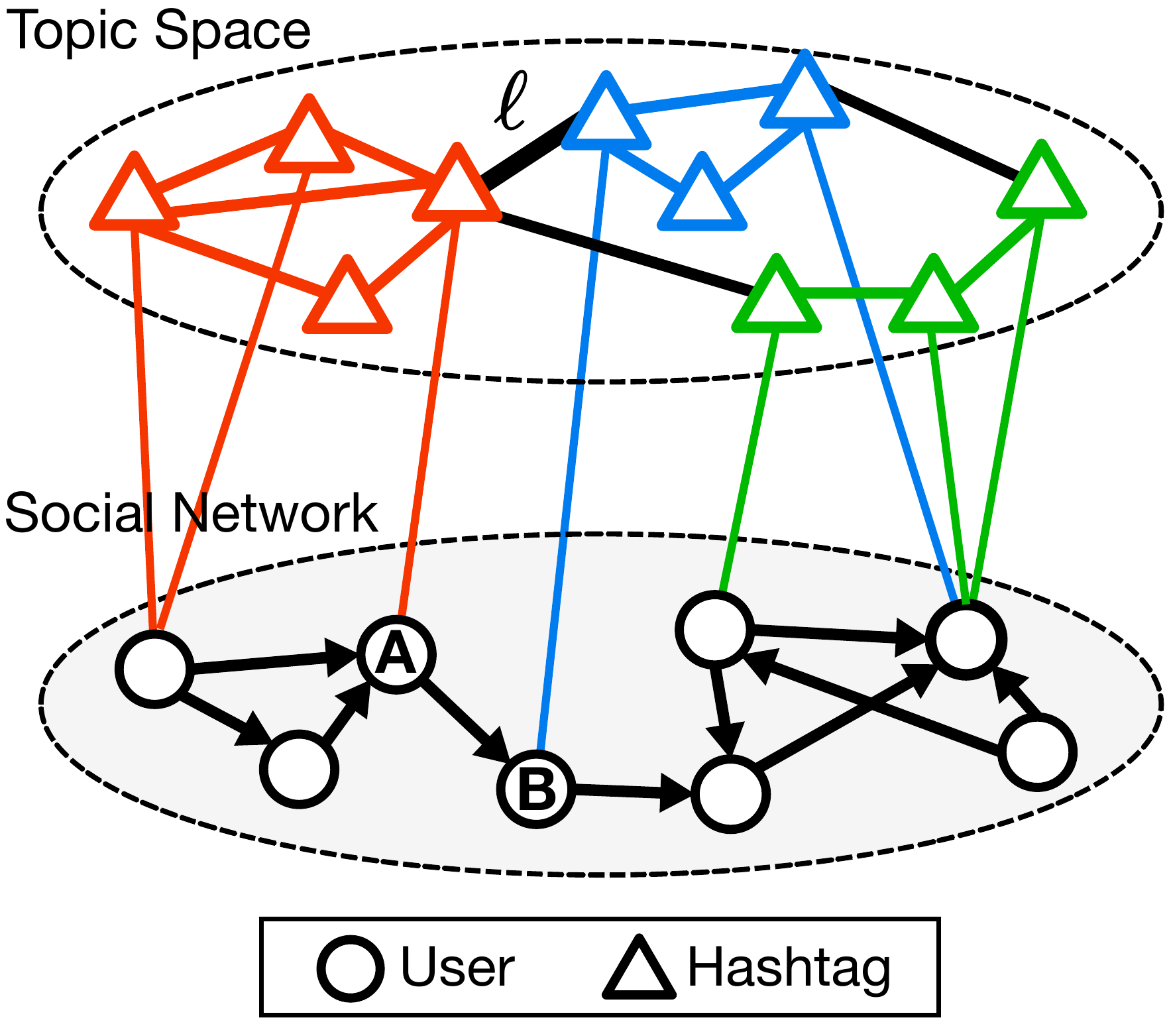}
\vspace{-1em}
\caption{We can represent the topics of online conversations in social media by a multi-layer network. The social network connects people. In the topic network, nodes represent hashtags that are linked when they co-occur; clusters represent topics (shown in colors). A person and a hashtag are connected when the person uses the hashtag.
}
\label{fig:topic_communities} 
\end{figure}

Online social media provide a platform on the Internet in which people can easily and cheaply  exchange messages. A great body of information is generated and tracked digitally, creating unprecedented opportunities for studying user generated content and information diffusion processes~\cite{Lazer2009,Vespignani2009}. 
Messages in social media involve a variety of \emph{topics}. Content, messages, or ideas are deemed semantically similar if they discuss, comment, or debate about the same topic; conversely, we can detect a topic by clustering a group of similar messages observed. 
In this paper we study the topical diversity of content and content creators --- messages and messengers. We develop a method to detect topics and propose a way to distinguish messengers with diverse interests from those with focused attention, as well as messages on general matters from those on particular domains. Thus we are able to tell which categories of users and messages have better chances to gain influence and impact.

In this study we use data from Twitter, one of the most popular social media platforms where Internet memes are generated, multiplied, and propagated. A user \emph{follows} others to subscribe to the information they share. This generates a social network structure along which messages spread. Each Twitter user can post short messages called \emph{tweets}, which may contain explicit topical tags, words, or phrases following a hash symbol (`\#'), named \emph{hashtags}. By using hashtags, people explicitly declare their interests in corresponding discussion and help others with similar preferences find appealing content.

The social network topology is determined by how people are connected. Each individual is represented as a node and each following relationship as an edge linking a pair of users (see the bottom layer in Fig.~\ref{fig:topic_communities}). Hashtags spread among people through these social connections and can be mapped into a semantic space, in which each node is a tag and similar ones are coupled forming topic clusters (see the top layer in Fig.~\ref{fig:topic_communities}). By examining which topics are attached to a user's messages, we can infer her interests; by examining the topics of tags that co-occur with a given hashtag, we can learn what that hashtag is about.
In reality we are able to observe the social network structure and information diffusion flows, but not topic formation in the semantic space. To the best of our knowledge, the connection between these two layers of information diffusion is not yet well explored~\cite{serrano2008self,romero2013interplay}. 

Let us highlight the main contributions of this paper:
\begin{itemize}

\item We develop a way to extract topics from online conversation. A network of hashtags is built by counting how many times a pair of hashtags appear together in a post. Communities (clusters of densely connected nodes) in such a network are found to well represent topics as sets of semantically related hashtags.

\item Given a user, we gauge the diversity of his topical interests by examining to which topic clusters each of his hashtags belongs. We can thus distinguish users with diverse interests from those with focused attention. The topical diversity of a hashtag is measured similarly, by considering its co-occurring hashtags. 

\item When a hashtag is adopted by people with diverse interests, or co-occurs with other tags on assorted themes, it is more likely for the tag to become popular. One interpretation is that diversity increases the probability of the hashtag of being exposed to different audience groups. We show that topical diversity of early adopters or co-occurring tags are good predictors for the future popularity of hashtags. 

\item In contrast, high topical diversity is not a helpful factor in the growth of individual social impact. Focusing on one or a few topics may be a sign of expertise. Inactive users attract followers by mentioning a variety of topics, while active users tend to obtain many followers by maintaining focused topical interests. Focused topical preferences promote the content appeal of ordinary users and celebrities alike.

\end{itemize}

\section{Background}

One prerequisite task for identifying topical interests of messages and messengers is to identify topics. Several studies examined the recognition of topics in the online scenario and social media~\cite{leskovec2009meme,xie2011visual,simmons2011memes,agarwal2012real,ferrara2013clustering}. Leskovec \emph{et al.}~\cite{leskovec2009meme} grouped short, distinctive phrases by single-rooted directed acyclic graphs used as signatures for different topics. Features extracted from content, metadata, network, and their combinations were leveraged to detect events in social streams~\cite{aggarwal2012event,ferrara2013clustering}. Another approach is based on the discovery of dense clusters in the inferred graph of correlated keywords, extracted from messages in a given time frame~\cite{agarwal2012real,tang2009social}. Here we adopt a similar strategy to identify clusters of similar hashtags by detecting communities in the network topology~\cite{blondel2008fast, Newman2012review} on account of topic locality.

Topic locality in the Web describes such a phenomenon that most Web pages tend to link with related content~\cite{Davison2000Topicality,Menczer2004cluster}. The effect of topic locality is utilized in focused Web crawlers~\cite{fil98crawler}, collaborative filtering~\cite{Goldberg1992CF,Goldberg2001}, interest discovery in social tagging~\cite{Schifanella20120folk,luca2012homophily}, and many other applications~\cite{haveliwala2003topic,michlmayr2007,tang2009social,weng2010twitterrank}. In our scenario, topic locality refers to the assumption that semantically similar hashtags are more likely to be mentioned in the same messages and therefore to be close to each other in the hashtag co-occurrence network.

We see a growing literature on discovering user interests and topics~\cite{java2007why,Phelan2009recommend,michelson2010discovering,chen2010short,weng2010twitterrank,xu2011discovering}. A common approach to use a vector representation generated from all the posts by a user to represent her interest. Then whether a user would be interested in a newly incoming message is determined by the similarity between feature vectors of user interests and the message~\cite{chen2010short,weng2012scirep}. LDA has also been applied to extract user interests from user generated content~\cite{weng2010twitterrank}. Java \emph{et al.}~\cite{java2007why} looked into communities of users in the reciprocal Twitter follower network and summarized user intent into several categories (daily chatter, conversations, information sharing, and news updates); a user could talk about various topics with friends in different communities. Michelson and Macskassy discovered entities mentioned in tweets according to predefined folksonomy-based categories to allocate topics so as to build an entity-based topic profile~\cite{michelson2010discovering}.
The diversity of user interests has not yet been thoroughly investigated. An exception is the work of An \emph{et al.}, who explored which news sources Twitter users are following and correlated the observation with the diversity of their political opinions~\cite{an2011media}. In this paper we propose a simple but powerful method to detect topics and infer user interests, as well as definitions of topical diversity of users and content.

Hashtag popularity has been examined from various perspectives, including their innate attractiveness~\cite{Berger:2009viral,Salganik2006music}, the network diffusion processes~\cite{daley64,goffman1964nature,vespignani2001SISonSF,aral2011viral,weng2013viral,weng2014prediction}, user behavior~\cite{weng2012scirep,ienco2010meme,Mei2012tags}, and the role of influentials along with their
adoption patterns~\cite{Kitsak2010kcore,bakshy2011everyone}. Romero \textit{et al.}~\cite{romero2013interplay} predicted popularity of a tag based on the social connections of its early adopters, but did not consider topicality and connections among tags.

We believe that the proposed measurement of topical diversity would prompt new approaches to the prediction of future hashtag popularity. Several previous studies have supported our intuition. For example, network diversity was shown to be positively correlated with regional economic development~\cite{quigley1998div,eagle2010div}; community diversity at the early stage tend to boost the chances of a meme going viral~\cite{weng2013viral,weng2014prediction}.

Many methods for quantifying social impact and identifying influential users have been proposed. User influence can be quantified in terms of high in-degree in the follower network~\cite{cha2010measuring,Suh2010}, information forwarding activity~\cite{romero2011influence,Suh2010},  seeding larger cascades~\cite{Kitsak2010kcore,bakshy2011everyone}, or topical similarity~\cite{tang2009social,weng2010twitterrank}.

\section{Methods}

In this section we describe our dataset and define several key concepts to facilitate the subsequent discussion.

\subsection{Dataset}

We collected public tweets from January to March 2013 using the Twitter public streaming API.\footnote{\url{http://dev.twitter.com/docs/streaming-apis}} We set the first two months as the \emph{observation period} and the last month as the \emph{test period}; the former is used to build up the topic network and quantify user topical interests, and the latter works for evaluating the results of prediction tasks. Table~\ref{table:dataset} shows several basic statistics about the dataset, which is publicly available at \url{carl.cs.indiana.edu/data/index.html#topic2014}.

\begin{table}
\caption{Basic statistics of the dataset, which is split into two periods: observation and testing. About 13\% of the tweets contain hashtags.}
\centering
\begin{tabular}{l r r}
\multirow{2}{*}{} & Jan-Feb 2013 & Mar 2013 \\
& (Observation)  & (Testing) \\
\hline
\# Tweets & 2,449,711,388 & 1,339,702,599 \\
\# Tweets with hashtags & 316,668,998 & 173,823,786 \\
\# Hashtags & 27,923,499 & 16,802,087 \\
\# Users  & 92,356,790 & 72,963,020\\
\hline
\end{tabular}
\label{table:dataset}
\end{table}

Hashtags during March 2013 are used for prediction tasks. We are interested in newly emergent tags, so that we are able to identify the start of their lifetime and track their growth for at least three weeks. We select hashtags that do not appear during January and February 2013, but are used by at least three distinct users during March 2013. In addition, only tags with the first tweet observed during the first week of March are considered, so that we can track their usage during the whole month. Eventually, 509,868 hashtags (3.03\% of all hashtags in March) were chosen as \emph{emergent} hashtags.

\subsection{Topic Clusters}

\begin{figure*}[t]
\centering
\includegraphics[width=\textwidth]{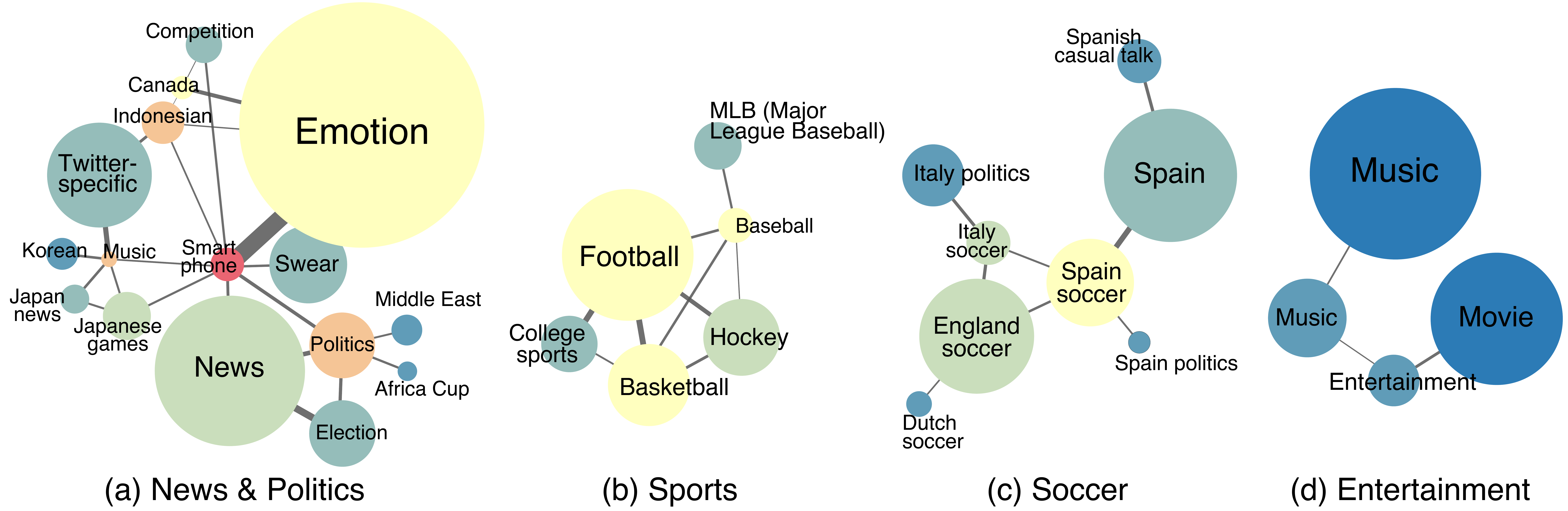}
\vspace{-2em}
\caption{Examples of connected topic clusters of related themes: (a) news and politics, (b) sports, (c) soccer, and (d) music and entertainment. Each node represents a cluster of hashtags on the topic as labelled; the area is proportional to the number of hashtags that the topic cluster contains; the color is assigned according to the degree so that high degree is more red and low degree is more blue. All these examples support the existence of topic locality.}
\label{fig:example_topics}
\end{figure*}

\begin{table}
\caption{Examples of topic clusters in the hashtag co-occurrence network.}
\centering
\begin{tabular}{l p{5.4cm}}
Semantics & Example \#hashtags\\
\hline
Technology & google, microsoft, supercomputers, ibm, wikipedia, pinterest, startuptip, topworkplace \\
Politics & tcot, p2, top, usgovernment, dems, owe, politics, teaparty \\
Lifestyle & pizza, pepsi, cheese, health, vacation, caribbean, ford, honda, volkswagen, hm, timberland \\
Twitter-specific & followme, followfriday, friday, justretweet, instantfollower, rt \\
Mobile devices & apple, galaxy3, note2, iosapp, mp3player, releases \\
\hline
\end{tabular}
\label{table:example_topics}
\end{table}

Hashtags are explicit topic identifiers on Twitter that are invented autonomously by millions of content generators. Since there is no predefined consensus on how to name a topic, multiple duplicate hashtags may be developed to represent the same event, theme, or object. For instance, \#followback, \#followfriday, \#ff, \#teamfollowback, and \#tfb are all about asking others to follow someone back or suggesting people to follow; \#tcot, \#ttxcot, \#twcot, and \#ccot label politically conservative groups on Twitter. 
To reduce the duplication, we shift attention from single hashtags to more general categories ---  clusters of semantically similar hashtags --- that we call \emph{topic clusters}. 

With the topic locality assumption that semantically similar hashtags are more likely to appear in the same tweets together, such topic clusters are expected to be densely connected. 
We detect these clusters by finding communities in the hashtag co-occurrence network. 
First we recover the network by only considering hashtags used by at least three distinct users and  join occurrences observed in at least three messages. We do this to filter out noise from accidental co-occurrence and spam. The recovered network contains 974,529 nodes and 7,325,492 edges. 
Then communities are detected using the Louvain community detection method~\cite{blondel2008fast}, which was selected because of its efficiency. We obtain 37,067 communities (the level 2 in the hierarchical structure found by the Louvain method). As exemplified in Table~\ref{table:example_topics}, communities in the hashtag co-occurrence network capture coherent topics. At the macroscopic level we can still observe strong topic locality (see Fig.~\ref{fig:example_topics}).

\subsection{Diversity of User Interests}
\label{sec:user_div}

Given a messenger $u$, we can track the sequence of hashtags (with repetition) that he used in the past, $h_1, h_2, \dots, h_{n_u}$. Each hashtag $h_i$ is attached to a topic $T(h_i)$, given by: 
\begin{equation}\label{formula:topic}
 T(h_i) =
  \begin{cases}
   C(h_i)	& \text{if $h_i$ exists in the hashtag} \\ 
   		& \text{co-occurrence network} \\
   h_i 		& \text{otherwise}
  \end{cases}
\end{equation}
where $C(h)$ is a community containing $h$ in the hashtag co-occurrence network. The set of distinct topics associated with all of $u$'s hashtags is denoted as $\mathbb{T}_u$, $T(h_i) \in \mathbb{T}_u$.
The topical diversity of a user's interests can be estimated by computing the entropy of hashtags across topics: 
\begin{eqnarray}
H_1(u) &=& -\sum_{T_j \in \mathbb{T}_u} P(T_j) \log P(T_j) \label{formula:div} \\
P(T_j) &=& \frac{1}{n_u} |\{h_i | T(h_i) = T_j, 1 \leq i \leq n_u\}|.
\end{eqnarray}

Table~\ref{table:example_users} compares two people, both having used 10 distinct hashtags for 20 times. User A was interested in trendy Twitter-specific tags almost exclusively (low $H_1$), while user B paid attention to a set of very diverse conversations about countries, movies, books, and horoscope (high $H_1$). Note that the opposite (and wrong!) conclusion, $H_1 > H_2$, would be drawn had we measured entropy based on hashtags rather than topic clusters.

\begin{table}
\caption{Comparison of two users with different diversity of topical interest.}
\centering
\begin{tabular}{c c p{6.1cm}}
User & $C$ & \#Hashtag (usage count) \\
\hline
\multirow{2}{*}{\textbf{A}} 
 & 20 & nowplaying(1) \\ 
 & 96 & rt(4), follow(3), tfb(2), ff(2), 500aday(2), teamfollow(2), teamfollowback(2), f4f(1), rt2gain(1) \\ 
 \cline{2-3}
 & \multicolumn{2}{r}{$n_A = 20$,~~$| \mathbb{T}_A |=10$,~~$H_1(A) = 0.2864$} \\
\hline
\multirow{2}{*}{\textbf{B}} 
 & 9 & australia(1) \\ 
 & 20 & cosmicconsciousness(1),thenotebook(1) \\ 
 & 33 & thedescendants(1) \\
 & 57 & friendswithbenefits(1) \\
 & 79 & thepowerofnow(2) \\
 & 139 & gemini(8), geminis(2) \\
 & 806 & tdl(2) \\ 
 & -- & tipfortheday (1) \\ 
 \cline{2-3}
 & \multicolumn{2}{r}{$n_B = 20$,~~$|\mathbb{T}_B|=10$,~~$H_1(B) = 2.3610$} \\
\hline
\end{tabular}
\label{table:example_users}
\end{table}

\subsection{Diversity of Content}

Similarly, given a hashtag $h$, we recover the sequence of other hashtags (with repetition) that co-occurred with it, $h_1, h_2, \dots, h_{m_h}$. Each co-occurring hashtag (\emph{co-tag}) $h_i$ is assigned to topic $T(h_i)$ based on the topic cluster to which it belongs (see Equation~\ref{formula:topic}). Then the co-tag diversity of $h$, $H_2(h)$, is measured in the same way as the user diversity $H_1$ (see Equation~\ref{formula:div}).

\section{Predicting Hashtag Popularity}

Do diversity measures help us detect hashtags that will go viral in the future? In this section we explore whether the topical diversity of a hashtag's adopters or co-tags predicts its future popularity.

\subsection{Prediction via User Diversity}
\label{sec:pred_user}

\begin{figure}[t]
\centerline{\includegraphics[width=0.85\columnwidth]{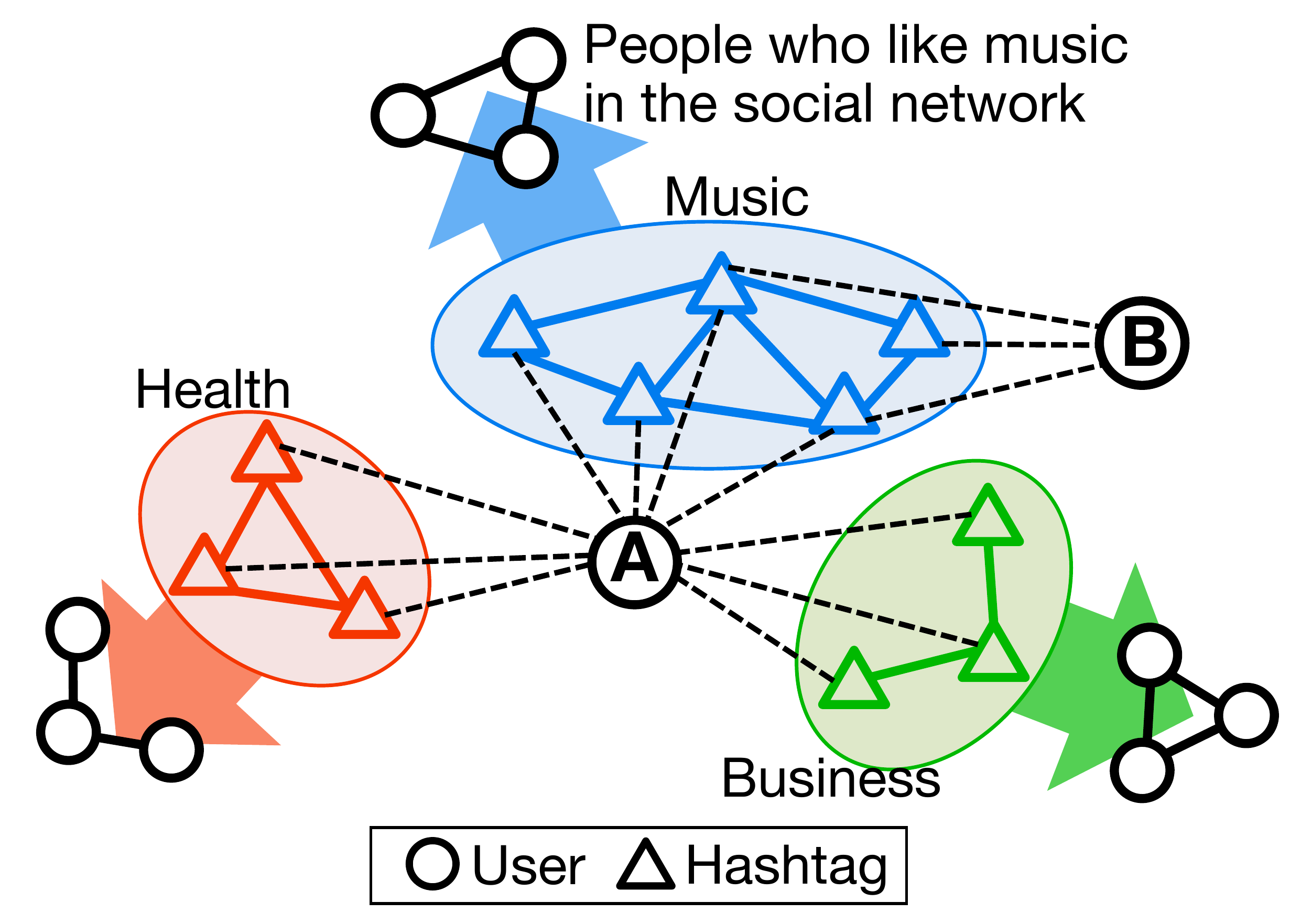}}
\vspace{-1em}
\caption{User A has diverse topical interests, each connected group corresponding to a social circle with common interests. User B displays more focused interests.}
\label{fig:user_div} 
\end{figure}

Hashtags in Twitter can be treated as channels connecting people with shared interests, because hashtags label and index messages enabling people to easily retrieve information and broadcast to certain groups. As illustrated in Fig.~\ref{fig:user_div}, users with focused interests are linked with few groups, while people who care about diverse issues are exposed to a larger number of interest groups through hashtag channels. 
We expect the latter category of users to play a critical bridging role, connecting many groups in the network. This would allow them to spread innovative information to multiple groups, as suggested by the weak tie hypothesis~\cite{Granovetter}, thus boosting the diffusion of hashtags~\cite{JP2007pnas,weng2013viral,weng2014prediction}. In other words, we hypothesize that if a hashtag has early adopters with diverse topical interests, it is more likely to go viral.

\begin{figure}
\centering
\includegraphics[width=\columnwidth]{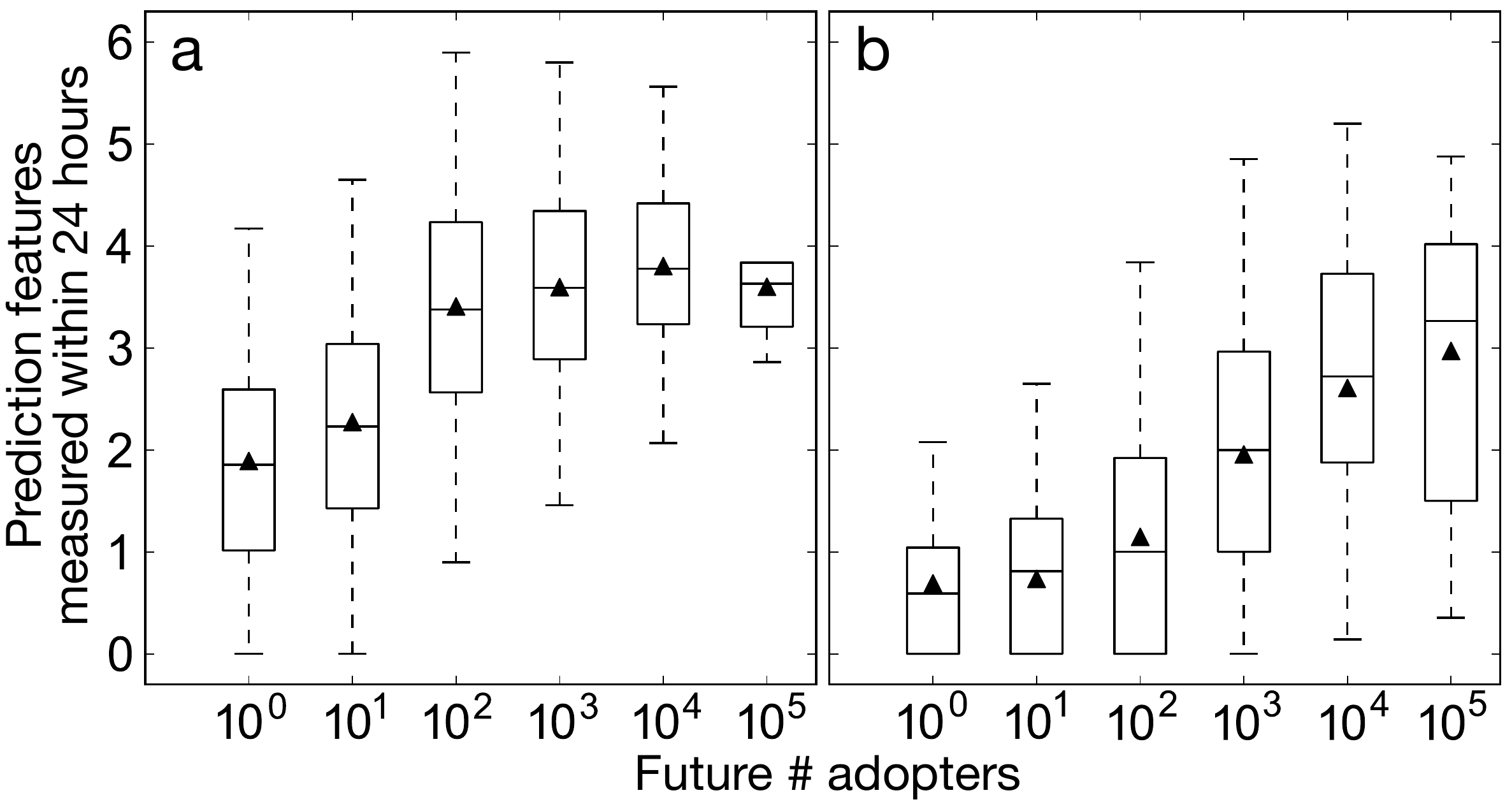}
\vspace{-2em}
\caption{(a) Correlation between the average topic-based entropy $H_1$ of adopters in the first 24 hours and the total number of future hashtag adopters. (b) Entropy $H_2$ of hashtags co-occurring in the first 24 hours across topic clusters as a function of future popularity of emergent hashtags.}
\label{fig:predict_corr}
\end{figure}

Given a hashtag $h$, we track the users who adopt it within $t$ hours after $h$ is created and compute the average interest diversity among these early adopters as a simple predictor. Irrespective of how long we track, we observe a positive correlation between the average user diversity and the future popularity of the hashtags, measured as the total number of adopters after one month (see Fig.~\ref{fig:predict_corr}a).

To better evaluate the predictive power of adopter diversity, let us run a simple prediction task based on information at the early stage to forecast which hashtags from the test period will be popular in the future. A hashtag is deemed popular if the number of distinct adopters at the end of the test period is above a given threshold. Our evaluation algorithm has three steps:
\begin{enumerate}[i)]

\item For each feature, we compute its value for each newly emergent hashtag $h$ in the test period based on the set of early adopters of $h$ within $t$ hours after the birth of $h$. A hashtag is born when the first tweet containing it appears. The feature is either a measure of user characteristics averaged among early adopters, or a linear combination of several such measures. We track adoption events for $t=1,6$, and 24 hours since birth.

\item Hashtags are ranked by the feature values in descending order.

\item We set a percentile threshold for labeling popular hashtags. The most popular hashtags are deemed ``viral.'' Based on this ground truth, we can measure false positive and true positive rates and draw a receiver-oper\-ating-character\-is\-tic (ROC) plot. The area under the ROC curve (AUC) is our evaluation metric. The higher the AUC value, the better the feature as a predictor of future hashtag popularity. 

\end{enumerate}

We consider several user attributes of early adopters that have been shown in the literature to be strong predictors of virality~\cite{cha2010measuring,Suh2010,Szabo2010pred,romero2011influence,weng2014prediction}. These include the number of early adopters $n$, number of followers \emph{fol} (potential audience), and number of tweets \emph{twt} that a user has produced during the observation period (activity). We additionally consider the diversity of topical interests, $H_1$. 
The goal of our experiment is not to achieve the highest accuracy (a task for which different learning algorithms could be explored). We aim to compare the predictive powers of different features. Therefore we focus on the relative differences between AUC values generated by single or combined features rather than on the absolute AUC values.
AUC values measured using different features are listed in Table~\ref{table:AUC_user_t}. Among individual features, $n$ is the most effective. When we combine it with other features, \emph{fol} yields high AUC consistently, but the differences are very small. The performance of the diversity metric is competitive, matching the top results in several experimental configurations. These results are not particularly sensitive to the popularity threshold or the duration of the early observation window.

\begin{table*}[t]
\caption{AUC of prediction results using different adopter features within $t$ early hours. Prediction features include the number of followers ($fol$), the number of tweets ($twt$), the diversity of topical interests of adopters ($H_1$), and the number of early adopters ($n$). The threshold is expressed as a top percentile of most popular hashtags that are deemed viral for evaluation purposes. Best results for each column are bolded.}
\centering

\begin{tabular}{r || ccc | ccc | ccc | ccc }
\hline
Threshold & \multicolumn{3}{c|}{50\%}& \multicolumn{3}{c|}{10\%}& \multicolumn{3}{c|}{1\%}& \multicolumn{3}{c}{0.1\%}\\
$t$ (hour)& 1 & 6 & 24 & 1 & 6 & 24 & 1 & 6 & 24 & 1 & 6 & 24 \\
\hline
$twt$
	& 0.57 & 0.58 & 0.59
	& 0.62 & 0.63 & 0.64
	& 0.67 & 0.68 & 0.68
	& 0.70 & 0.71 & 0.72
\\
$fol$
	& 0.57 & 0.58 & 0.59
	& 0.62 & 0.64 & 0.65
	& 0.67 & 0.69 & 0.70
	& 0.74 & 0.75 & 0.76
\\
$H_1$
	& 0.55 & 0.55 & 0.55
	& 0.58 & 0.58 & 0.58
	& 0.60 & 0.60 & 0.61
	& 0.63 & 0.63 & 0.64
\\
$n$
	& 0.64 &  \textbf{0.68} & 0.71
	& 0.75 & 0.79 & 0.84
	& 0.82 & 0.86 &  \textbf{0.90}
	& 0.86 & \textbf{0.89} & 0.91
\\
\hline
$n+twt^\dagger$
	& 0.64 & \textbf{0.68} & 0.71
	& 0.75 & \textbf{0.80} & 0.84
	& \textbf{0.83} & \textbf{0.87} & \textbf{0.90}
	& \textbf{0.87} & \textbf{0.89} & \textbf{0.92}
\\
$n+fol^\dagger$
	& \textbf{0.65} & \textbf{0.68} & \textbf{0.72}
	& \textbf{0.76} & \textbf{0.80} & \textbf{0.85}
	& \textbf{0.83} & \textbf{0.87} &  \textbf{0.90}
	& \textbf{0.87} & \textbf{0.89} & \textbf{0.92}
\\
$n+H_1^\dagger$
	& 0.64 & \textbf{0.68} & 0.71
	& 0.75 & \textbf{0.80} & \textbf{0.85}
	& \textbf{0.83} & 0.86 &  \textbf{0.90}
	& 0.86 & \textbf{0.89} & \textbf{0.92}
\\
\hline
\multicolumn{13}{l}{$\dagger$~A linear combination with coefficients determined by regression fitting using least squared error.}
\end{tabular}

\label{table:AUC_user_t}
\end{table*}

\subsection{Prediction via Content Diversity}
\label{sec:pred_cotag}

\begin{table*}[t]
\caption{AUC of prediction results using different features among co-tags within $t$ early hours. Prediction features include the number of tweets containing the co-tags ($T$), the number of co-tag adopters ($A$), the diversity of co-tags ($H_2$), and the number of observed co-tags ($m$). The threshold is expressed as a top percentile of most popular hashtags that are deemed viral for evaluation purposes. Best results for each column are bolded.}
\centering
\begin{tabular}{r || ccc | ccc | ccc | ccc }
\hline
Threshold & \multicolumn{3}{c|}{50\%}& \multicolumn{3}{c|}{10\%}& \multicolumn{3}{c|}{1\%}& \multicolumn{3}{c}{0.1\%}\\
$t$ (hour)& 1 & 6 & 24 & 1 & 6 & 24 & 1 & 6 & 24 & 1 & 6 & 24 \\
\hline
$T$
	& 0.50 & 0.51 & 0.52
	& 0.51 & 0.53 & 0.55
	& 0.58 & 0.62 & 0.66
	& 0.66 & 0.72 & 0.75
\\
$A$
	& 0.50 & 0.50 & 0.52
	& 0.50 & 0.52 & 0.54
	& 0.58 & 0.62 & 0.65
	& 0.65 & 0.71 & 0.74
\\
$H_2$
	& 0.50 & 0.51 & 0.53
	& 0.52 & 0.54 & 0.57
	& 0.61 & 0.66 & 0.70
	& 0.70 & 0.77 & 0.82
\\
$m$
	& 0.52 & 0.53 & 0.55
	& 0.55 & 0.58 & 0.61
	& 0.64 & \textbf{0.70} & \textbf{0.75}
	& 0.72 & \textbf{0.81} & \textbf{0.86}
\\
\hline
$m+T^\dagger$
	& 0.52 & 0.53 & 0.55
	& 0.54 & 0.57 & 0.61
	& 0.64 & \textbf{0.70} & \textbf{0.75}
	& 0.72 & \textbf{0.81} & \textbf{0.86}
\\
$m+A^\dagger$
	& 0.52 & 0.53 & 0.55
	& 0.55 & 0.57 & 0.61
	& 0.64 & \textbf{0.70} & \textbf{0.75}
	& 0.72 & \textbf{0.81} & \textbf{0.86}
\\
$m+H_2^\dagger$
	& \textbf{0.55} & \textbf{0.55} & \textbf{0.57}
	& \textbf{0.58} & \textbf{0.60} & \textbf{0.63}
	& \textbf{0.66} & \textbf{0.70} & \textbf{0.75}
	& \textbf{0.74} & \textbf{0.81} & \textbf{0.86}
\\
\hline
\multicolumn{13}{l}{$\dagger$~A linear combination with coefficients determined by regression fitting using least squared error.}
\end{tabular}
\label{table:AUC_cotag_t}
\end{table*}

In this section we examine whether the future popularity of a hashtag is affected by the topical diversity of its early co-occurring tags.
How people apply hashtags to label their messages depicts their topical interests and determines the topology of the tag co-occurrence network. In Fig.~\ref{fig:topic_communities}, a link between the topic layer and the social layer of the network marks an association between a user and a hashtag. This tag may attract an audience in the social network. The co-occurrence of two tags extends the audience groups of both. For example, link $\ell$ in Fig.~\ref{fig:topic_communities} exposes user A to the blue topic and user B to the red cluster. 
Therefore we expect a hashtag to be exposed to more potential adopters, making it more likely to go viral, if it often co-occurs with many other hashtags. To test this hypothesis, we measure the number $m$ of co-tags. Furthermore, if co-tags are very popular, we would expect a stronger effect because they would provide a larger audience. We therefore measure the popularity of co-tags in terms of numbers of tweets $T$ and adopters $A$ during the observation period. And if co-tags are about diverse topics, this may further boost the effect by extending the audience to many groups with small overlap. In conclusion, we hypothesize that many popular co-tags about diverse topics should be a sign that a hashtag will grow popular.

Given an emergent hashtag $h$, we track other tags that co-occur with $h$ within $t$ hours after $h$ is born and measure the topical diversity $H_2$ of these co-tags. We observe a positive correlation between the diversity of early co-tags and the future popularity of the tag (see Fig.~\ref{fig:predict_corr}b). Then we apply the same method as in Sec.~\ref{sec:pred_user} to test the predictive power of different traits associated with early co-tags. In this case, the prediction features for each target hashtag are computed based on early co-tags instead of adopters. Again, the goal of our experiment is to compare the predictive powers of different features, thus we examine the relative differences in AUC values generated by the various traits. The results are reported in Table~\ref{table:AUC_cotag_t}. The number $m$ of co-tags observed in the early stage is the best single predictor of virality. When we combine $m$ with a second feature, co-tag diversity provides the best results  irrespective of the threshold or the duration of the early observation window. Interestingly, $m$ and $H_2$ are both about diversity and perform better than the popularity-based features $T$ and $A$.

\subsection{Summary}

In the discussion above, we evaluate the predictive powers of two categories of features for identifying future popular hashtags. These two sets of features, based on early adopters and co-tags, have different effectiveness. By comparing the AUC values in Tables \ref{table:AUC_user_t} and \ref{table:AUC_cotag_t}, we find that adopter features yield better results. However, they also require additional prerequisite knowledge: in addition to tracking hashtag co-occurrences for building the topic network, we also need to record user-generated content. The features built upon early co-tags are less expensive, but the performance is slightly worse; a possible interpretation for this is that few tweets in the observation window may contain co-occurring tags, while they all have associated users. Therefore co-tag features are more sparse.  
Depending on what type of information is available, one might choose either approach or a combination of both. 

\section{Social Influence}
\label{ref:user_popu}

High topical diversity of adopters and co-occurring tags is a positive sign that a hashtag is growing popular, as shown in the previous section. However, does high topical diversity also signal a growth in individual influence?
On one hand, when an individual talks about various topics, she may have contact with many others through shared interests or hashtags, thus attracting more attention (see Fig.~\ref{fig:user_div}).
On the other hand, focused interest may enhance expertise in specific fields, thus increasing the content interestingness and retweetability. In this light, low diversity triggered by expertise might help people become popular.
In this section we evaluate these two contradictory hypotheses.

Some people are more influential than others in persuading friends to adopt an idea, an action, or a piece of information. 
The concept of \emph{social influence} has been discussed extensively in social media research. Most of the studies in the literature have considered users who are active~\cite{cha2010measuring}, have many followers~\cite{cha2010measuring,romero2011influence}, are able to trigger large cascades~\cite{Kitsak2010kcore,bakshy2011everyone}, or get retweeted or mentioned a lot~\cite{cha2010measuring,Suh2010,romero2011influence} as signals of high social influence.
Which user characteristics make people popular and influential? Does the diversity of individual topical interests play a role in the social influence processes? Let us consider several individual properties:
\begin{description}
\item[Number of retweets ($RT$)] How many times an individual is retweeted during the observation time period. We consider $RT$ as a direct indicator of social influence, since it quantifies how many times the user succeeds in making others adopt and spread information.\footnote{Due to the settings of the Twitter API, the number of retweets per user that we collect includes all the retweeters in every cascade. That is, suppose user B retweets user A and then C retweets B; both tweets are counted in $RT$ for A, even though C did not directly retweet A. However, since the majority of information cascades are very shallow~\cite{bakshy2011everyone}, $RT$ is a good approximation of the direct retweet count.} The number of retweets is dependent on the length of the observation window, because we believe that social influence is accumulated in time and requires long-term endeavor~\cite{cha2010measuring}.

\item[Number of followers ($fol$)] The number of followers suggests how many people can potentially view a message once the user posts it.

\item[Number of tweets ($twt$)] The number of tweets generated by the user; the higher the number, the more active the user.

\item[Content interestingness ($\beta$)] How interesting is the content posted by the user. Lerman studied the interestingness of online content on Digg and defined it as ``the probability it will get retweeted when viewed''~\cite{lerman2007social}. To measure $\beta$ in the Twitter context, we assume that the value of $RT$ for an individual is proportional to the number of tweets $twt$ he produced, the number of followers $fol$, the chance $\alpha$ that a message is seen by a follower, and the appeal of the content. Treating $\alpha$ as a constant for simplicity, we obtain
\begin{equation}
\beta = \frac{RT}{twt \cdot fol \cdot \alpha} \propto \frac{RT}{twt \cdot fol}.
\end{equation}

\item[Diversity of interests ($H_1$)] See Sec.~\ref{sec:user_div}.

\end{description}

%
%
\begin{table}
\caption{Linear regression estimating how many times a user is retweeted. For efficiency, the regression is based on a random sample of 10\% of the users ($N = 2,171,624$).} 
\centering
\begin{tabular}{r >{\hfill}p{2cm} >{\hfill}p{1.4cm}}
\hline
& Coefficient & SE \\
\hline
(Intercept) & 20.9~~~~~~~ & 0.5 \\
Num. followers ($fol$)$^\dagger$ & 193.0 *** & 0.5 \\
Num. tweets ($twt$)$^\dagger$ & 51.1 *** & 0.5 \\
Content interestingness ($\beta$)$^\dagger$ & 3.9 *** & 0.5 \\
Diversity of interests ($H_1$)$^\dagger$ & -9.1 *** & 0.5 \\
\hline
\multicolumn{3}{l}{$\dagger$~Variables are normalized by $Z$-score. ***~$p < 0.001$}
\end{tabular}
\label{table:regression}
\end{table}

Table~\ref{table:regression} lists the results of a linear regression estimating how many times a user is retweeted according to several user features. 
Intuitively, users with many followers are more likely to spread their messages and thus get retweeted more frequently, because they have many more potential viewers. The number of followers is the most important factor, as supported by the largest positive coefficient in the regression.
The number of generated tweets also has a positive coefficient in the regression, implying that being active helps users get retweeted more. The result confirms several existing studies suggesting that high social influence requires long-term, consistent effort~\cite{cha2010measuring,Suh2010}.
The interestingness of the story is positively correlated with social influence as well, although not as strongly as the other factors.
Finally, the negative coefficient of diversity in Table~\ref{table:regression} suggests that users with diverse interests tend to have low influence. This supports the hypothesis that social influence is topic-sensitive, requiring expertise in a specific field~\cite{weng2010twitterrank}; posting about the same topic is more effective for gaining social influence, compared to commenting on many different subjects. 
In summary, people can acquire social influence by having a big audience group, being productive, creating interesting content, and staying focused on a field. Unfortunately, it seems that there is no simple recipe of success.

\begin{figure}[t]
\centering
\includegraphics[width=\columnwidth]{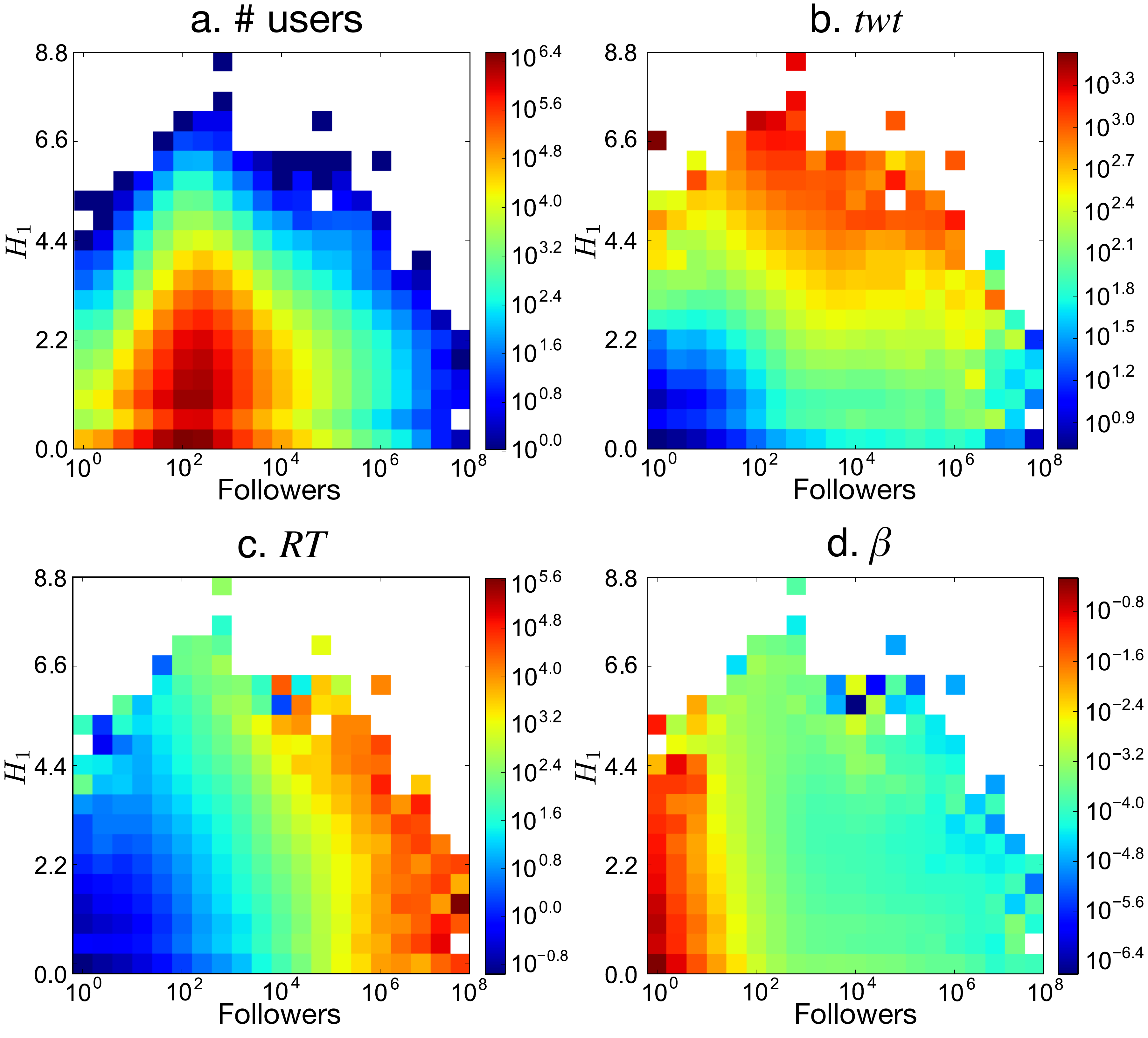}
\vspace{-2em}
\caption{Heatmaps of (a) the number of users, (b) the number of tweets generated, (c) how many times a user is retweeted, and (d) the content interestingness of a user, as a function of the diversity of topical interests, $H_1$, and the number of followers in the observation window. 
}
\label{fig:heatmaps}
\end{figure}

We illustrate how several user properties are related to the number of followers and the topical diversity of user interests in Fig.~\ref{fig:heatmaps}. Most users have a small number of followers and low entropy (Fig.~\ref{fig:heatmaps}a). Active users tend to have high diversity, as expected by the nature of entropy (Fig.~\ref{fig:heatmaps}b). The number of followers is shown in Fig.~\ref{fig:heatmaps}c to be a powerful factor to get retweeted more often, consistently with the regression results in Table~\ref{table:regression}. Finally, the content interestingness appears to be correlated with the number of followers but  strongly with user diversity (Fig.~\ref{fig:heatmaps}d).


\subsection{Active vs. Inactive Users}

\begin{figure}[t]
\centering
\includegraphics[width=\columnwidth]{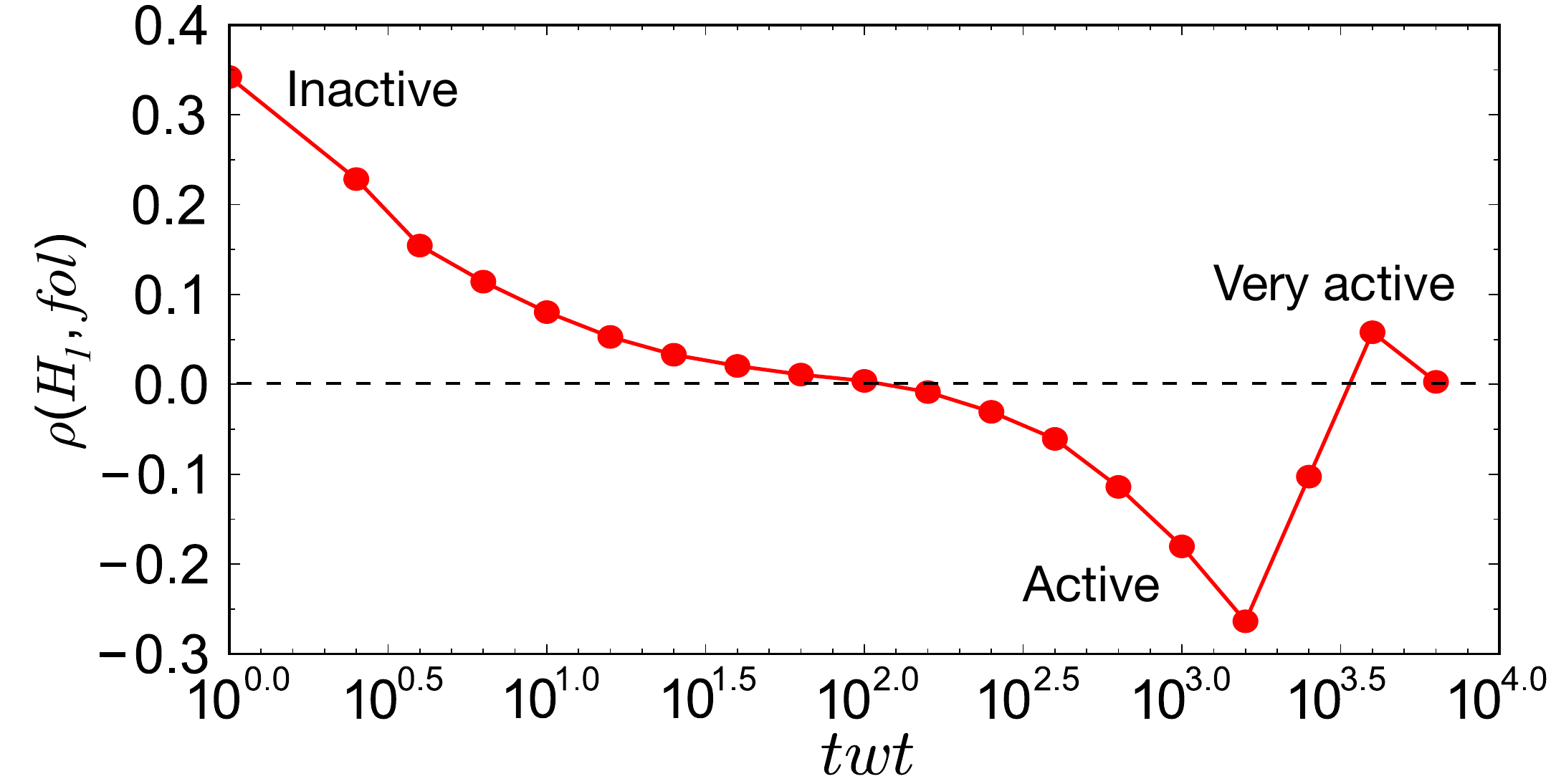}
\caption{Spearman rank correlation between the number of followers and the topical diversity of user interests as a function user activity. All shown correlation values are significant ($p<0.05$).
}
\label{fig:corr_H_fol}
\end{figure}

Let us explore how the number of followers a user can attract is affected by the diversity of topical interests. The entropy measure for diversity is biased by user activity: generating more tweets with more hashtags tends to yield higher entropy. Thus we group users by productivity, so that individuals in the same group have comparable values of topical diversity. For users in the same group, we compute the Spearman rank correlation between the number of followers and diversity. We use Spearman because, unlike Pearson, it does not require that both variables be normally distributed.
According to Fig.~\ref{fig:corr_H_fol}, low-engagement users attract followers by talking about various topics, while active users tend to obtain many followers by maintaining focused topical interests. For the most active users, topical diversity is not relevant; many of these accounts are spammers and bots.

\subsection{Celebrities and Ordinary Users}

\begin{figure}
\centering
\includegraphics[width=\columnwidth]{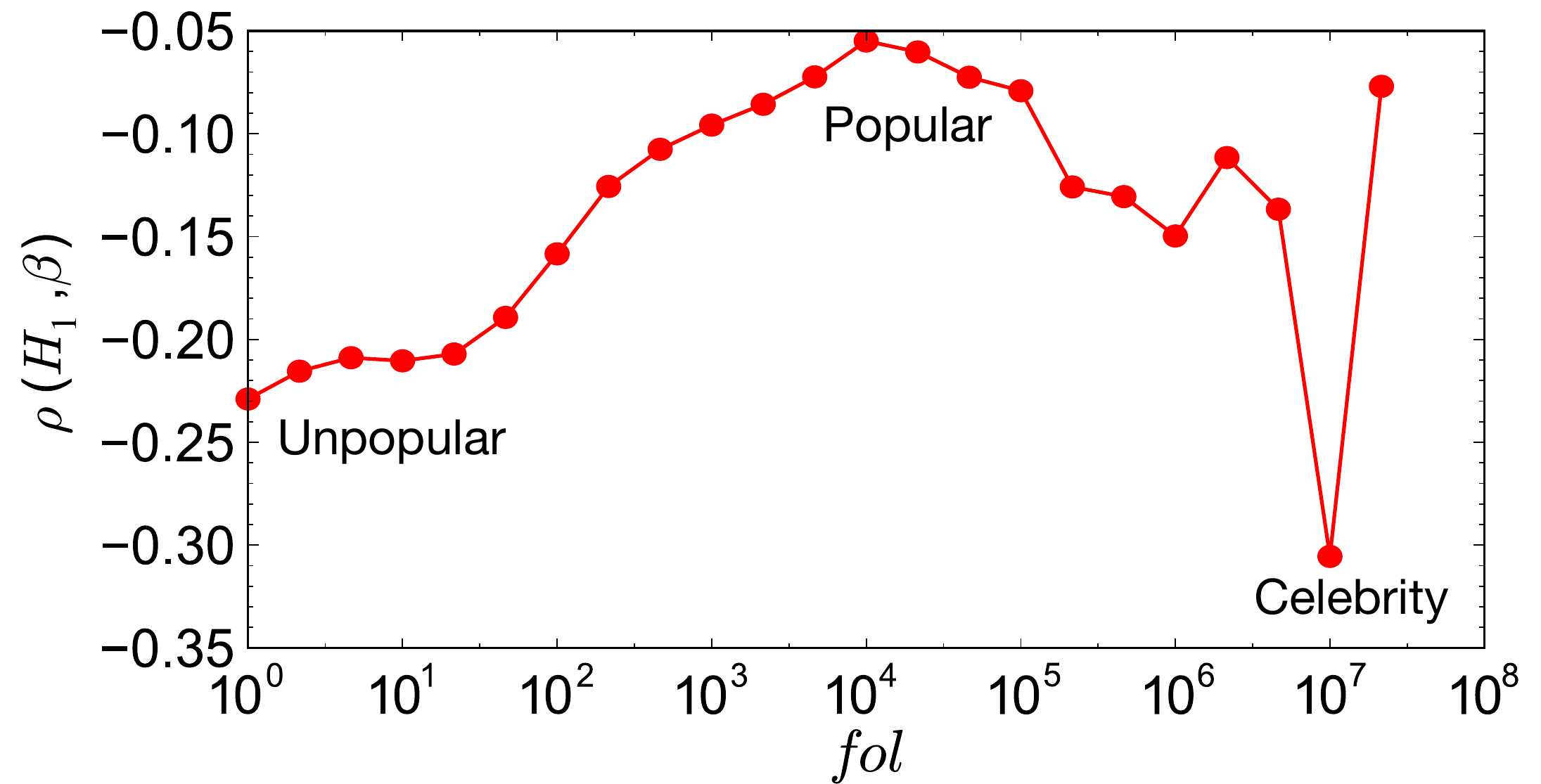}
\caption{Spearman rank correlation between content interestingness and the topical diversity of user interests as a function of how many followers users have. All shown correlation values are significant ($p<0.05$).
}
\label{fig:corr_H_beta}
\end{figure}

When looking into the effect of interest diversity on content appeal, we need to control for the number of followers, since our interestingness measure is strongly correlated with the number of followers (see Fig.~\ref{fig:heatmaps}d).
The negative correlations shown in Fig.~\ref{fig:corr_H_beta} suggest that in general, focused posts promote content appeal. One possible interpretation is that people follow someone for a reason. Content has to be consistent in order to match such expectations; i.e., one is less likely to share a tip on cosmetics from a politician.
This effect is stronger for users with few followers and celebrities; people with moderate popularity  generate retweets with focused and diverse content.

\section{Conclusion}

We proposed methods to identify topics using Twitter data by detecting communities in the hashtag co-occurrence network, and to quantify the topical diversity of user interests and content, defined by how tags are distributed across different topic clusters. 
We found that popular hashtags tend to have adopters who care about various issues and to co-occur with other tags of diverse themes at the early stage. One practical application evaluated in this paper is to predict viral hashtags using features built upon the topical diversity of early adopters or co-tags. 
%
In the prediction using information on early adopters, the performance of topical diversity is competitive  with other user features while combined with the number of early adopters. In the prediction with early co-occurring hashtags, features about diversity, including the number of early co-tags and their topical diversity, excel the popularity-based features.
%
However, high topical diversity is not a positive factor for individual popularity. High social influence is more easily obtained by having a big audience group, producing lots of interesting content, and staying focused. 
In short, diverse messages and focused messengers are more likely to generate impact.

The interesting observation that high diversity helps a hashtag grow popular but does not help develop personal authority originates from the different mechanisms by which a hashtag and a user attract attention. In the diffusion process of a hashtag, adopters with diverse interests play a role as bridges connecting different groups and thus positively improve the visibility of the tag. These results are consistent with Granovetter's theory~\cite{Granovetter}, as well as our recent findings on the strong link between community diversity and virality~\cite{weng2013viral}. On the other hand, a user gains social influence through expertise  or authority within a cohesive group with common interests.

Topical diversity provides a simple yet powerful way to connect the social network topology with the semantic space extracted from online conversation. We believe that it holds great potential in  applications such as predicting viral hashtags and helping users strengthen their online presence. 


\subsection*{Acknowledgement}

We are grateful to members of NaN (\url{cnets.indiana.edu/groups/nan}) for many helpful discussions. 
The work is supported in part by the James S. McDonnell Foundation, NSF grant CCF-1101743, and DARPA grant W911NF-12-1-0037.

\bibliographystyle{abbrv}
\bibliography{../finalrefs}

\end{document}